\documentclass[pra,twocolumn,superscriptaddress,floatfix,amsmath,nofootinbib,amssymb]{revtex4-1}
\usepackage[english]{babel}
\usepackage{graphicx}
\usepackage{dcolumn}
\usepackage{bm}
\usepackage{verbatim}
\usepackage{mathrsfs}
\usepackage{color}
\newcommand{\ket}[1]{\left| {#1} \right\rangle}

\newcommand{\ro}{r_{\omega}}
\newcommand{\eq}[1]{(\ref{#1})}

\newcommand{\up}{\uparrow}
\newcommand{\down}{\downarrow}
\newcommand{\qr}{q_\text{R}}
\newcommand{\ql}{q_\text{L}}

\begin{document}

\title{Entanglement of arbitrary spin fields in non-inertial frames}
\author{Miguel Montero}
\affiliation{Instituto de F\'{i}sica Fundamental, CSIC, Serrano 113-B, 28006 Madrid, Spain}
\author{Eduardo Mart\'{i}n-Mart\'{i}nez}
\affiliation{Instituto de F\'{i}sica Fundamental, CSIC, Serrano 113-B, 28006 Madrid, Spain}
\begin{abstract}
We generalise the study of fermionic and bosonic entanglement in non-inertial frames to fields of arbitrary spin and beyond the single mode approximation. After the general analysis we particularise for two interesting cases: entanglement between an inertial and an accelerated observer for massless fields of spin  $1$ (electromagnetic)  and  spin $3/2$ (Rarita-Schwinger). We show that, in the limit of infinite acceleration, no significant differences appear between the different spin fields for the states considered.
 \end{abstract}

\maketitle

\section{Introduction}\label{intro}

The novel field of relativistic quantum information has experienced a quick development in the recent past  \cite{Alsingtelep,TeraUeda2,ShiYu,AlsingMcmhMil,Alicefalls,AlsingSchul,caball,Adeschul,KBr,ManSchullBlack,PanBlackHoles,DH,Steeg,Edu2,schacross,Ditta,Hu,DiracDiscord,Edu6, Edu7,Edu8,Edu9,Edu10,Mig1}. Among other topics, this field includes the study of quantum correlations affected by gravitational effects or field state described by a non-inertial observer. It has been recently shown in \cite{Edu9} that the so-called single mode approximation \cite{Alsingtelep,AlsingMcmhMil} was misunderstood and, furthermore, does not hold in most of the cases. It was also shown that to properly take into account all the features of entanglement in non-inertial frames it is necessary to go beyond such an approximation \cite{Edu9,Edu10,Mig1,Mig2}.

So far, most of the works have only considered spinless fields, either bosonic or fermionic \cite{Alicefalls,AlsingSchul,caball,Adeschul,KBr,ManSchullBlack,PanBlackHoles,DH,Steeg,Edu2,schacross,Ditta,Hu,DiracDiscord,Edu9}. Only a few works have considered fields of non-zero spin in this context, only in very specific cases (spin 1/2) \cite{Edu2,Edu3,Edu4}, and always assuming the single mode approximation. In this work we provide the tools necessary to extend these studies to fields of arbitrary spin and beyond the single mode approximation.  We  do so via the explicit computation of the general expression for the vacuum and Unruh excitations in the Rindler basis for the arbitrary spin case. Given these expressions, the study of entanglement in any setting in which only a finite number of relevant modes play a role becomes straightforward. To illustrate this, we explicitly study entanglement behaviour as a function of acceleration for the particular case of fields of spin 1 and spin 3/2, cases that have not been properly studied before (see section \ref{mcPKing}).

\begin{figure}[hbtp] \centering
\includegraphics[width=.50\textwidth]{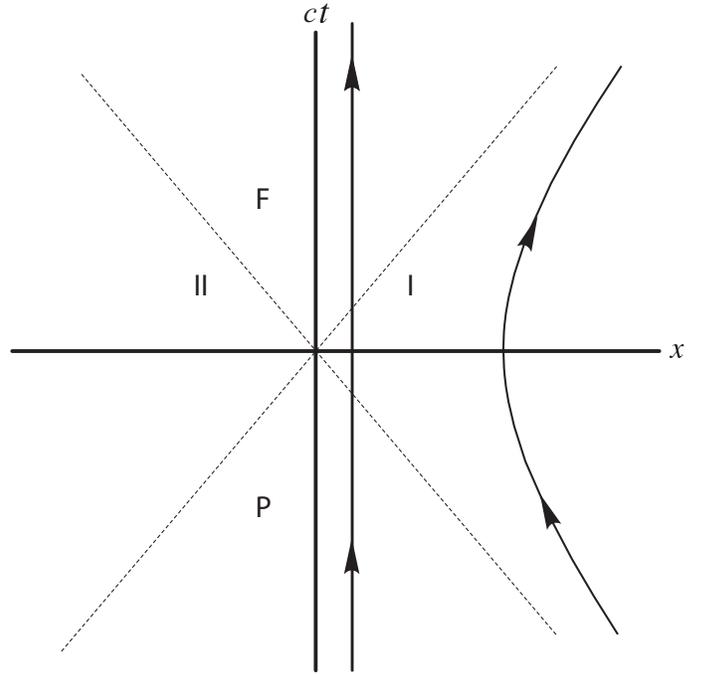}
\caption{Spacetime diagram, showing the trajectories of an inertial and an accelerated observer.}
\label{scheme}
\end{figure}

In our setting, and for the sake of simplicity, we shall consider a (1+1) dimensional spacetime, although the results can be readily extended to higher-dimensional spacetimes as well. Spin quantisation axis is chosen along the acceleration direction so no Thomas precession occurs, as it is common in relativistic quantum information literature \cite{AlsingSchul,Edu2,Edu9}. Throughout our work, we will refer to the causally disconnected left and right wedges of the flat spacetime shown in Fig. \ref{scheme} as regions I and II. The worldline of a uniformly accelerated Rindler observer must lie in either region I or II. Since both regions are globally hyperbolic, they admit independent quantum field theory constructions \cite{Wald2,Takagi}, each having its own set of creation and annihilation operators. If we want to build a quantum field theory for all of Minkowski spacetime, both of these constructions have to be taken into account, and therefore the total Hilbert space factorises as $\mathcal{H}_\text{I}\otimes \mathcal{H}_\text{II}$.  As it can be seen elsewhere \cite{Alicefalls,AlsingSchul,Edu2,Edu9}, entanglement effects in non-inertial frames are, in fact, related to the nontrivial change from the Minkowski to the Rindler basis.

We will first construct the fermionic inertial modes, and find expressions for the Minkowski vacuum and excitations in Rindler coordinates for arbitrary spin. This constitutes section \ref{ferm}. In section \ref{bos} we present the extension of the formalism to arbitrary spin bosonic fields. In section \ref{entsec} we study entanglement for some interesting states in the case of the electromagnetic and spin 3/2 fields. Finally, section \ref{cojoniak} contains our conclusions.

\section{Fermionic fields}\label{ferm}

In the context of fermionic fields, we can define a set of inertial modes that are expressed as monochromatic modes in the accelerated observer Fock basis. These modes are named `Unruh modes' \cite{Edu9}, and the creation operators associated with them are defined by
\begin{align}C^\dagger_{\omega,\sigma,\text{R}}&=\cos r_\omega c^\dagger_{\omega,\sigma,\text{I}}- \sin r_\omega d_{\omega,-\sigma,\text{II}},\nonumber\\
C^\dagger_{\omega,\sigma,\text{L}}&=\cos r_\omega c^\dagger_{\omega,\sigma,\text{II}}- \sin r_\omega d_{\omega,-\sigma,\text{I}},\label{umodes}\end{align}
Here, the operator $c_{\omega,\sigma,\text{I}}$ corresponds to the Rindler mode of frequency $\omega$ and spin $\sigma$ in region I, and $d_{\omega,\sigma,\text{I}}$ corresponds to its antiparticle, the same considerations applying to region II. The parameter $r_\omega$ is defined by
\begin{equation}\label{casiobomber}
\tan r_\omega=e^{-\pi\omega c/a},
\end{equation}
 where $a$ is the proper acceleration of the observer. Notice that the extension to massive fields is direct if we replace $\omega /c$ by $|\bm k|$ in \eqref{casiobomber} (See \cite{Jauregui, Edu9}).
 
It can be easily proved that regardless of the formal differences among inner products for different spin fields, equation \eq{umodes} is valid for arbitrary spin following the analytical continuation arguments in \cite{Wald2,ch1,Edu9} that also apply here. Analog expressions apply for the bosonic case (see section \ref{bos} and \cite{Takagi}). 

As \eq{umodes} shows, there are two distinct kinds of Unruh modes, which we label as right (R) and left (L) modes. A general Unruh mode is therefore a linear combination of the form
\begin{align}C^\dagger_{\omega,\sigma,\text{U}}=\qr C^\dagger_{\omega,\sigma,\text{R}}+\ql C^\dagger_{\omega,-\sigma,\text{L}}\label{um}\end{align}
satisfying the obvious normalisation condition $\vert\qr\vert^2+\vert\ql\vert^2=1$. The single mode approximation consisted in the assumption that the Unruh mode with $\qr=1$ is a good approximation for a Minkowski monochromatic mode. This is not the case, as such modes, when expressed in terms of Unruh modes, have important contributions from modes  \eq{um} with $\qr\neq1$ \cite{Edu9}. Therefore considering arbitrary Unruh modes is necessary in general. 

As shown elsewhere \cite{Wald2,Birrell}, the Minkowski vacuum can be factorised as a product of the vacua of all different Unruh modes,
\begin{align}\ket{0}_\text{M}=\bigotimes_{\omega}\ket{0}_{\omega,\text{U}}\label{product}\end{align}
where $\omega$ is the Rindler frequency associated to the Unruh mode (See, among others  \cite{Takagi,Edu9}). This means that each independent Unruh mode of Rindler frequency $\omega$ can be studied separately.

In order to express the Minkowski vacuum state in terms of Rindler modes we take advantage of the fact that the Minkowski vacuum is annihilated by all the Unruh annihilation operators, that is $C_{\omega,\sigma,\text{U}}\ket{0}_\text{M}=0$. Since we shall work only with Unruh modes of a single Rindler frequency, we may drop the label $\omega$ for the rest of the section. In other words, we only need to consider a single frequency sector of the vacuum state $\ket{0}_{\omega,\text{U}}$.

Although the condition $C_{\sigma,\text{U}}\ket{0}_\text{U}=0\ \forall \sigma$ uniquely determines the vacuum state, we still have to specify a Hilbert space basis. We will employ a number basis obtained by applying Rindler creation operators on the Minkowski vacuum, as it is commonplace in the field. Nevertheless, due to the fermionic nature of the field, we also have to specify the order in which the operators will act so as to completely specify the basis. The differences between these bases may have nontrivial effects on entanglement, a phenomenon thoroughly studied in \cite{Mig2}.

We will find that a specific fermionic operator ordering results particularly useful to generalise the results for arbitrary spin, keeping in mind that changing to any other ordering is trivial once the state has been computed.

 Before we obtain the expressions for the vacuum and arbitrary excitations for fermionic fields, we will introduce some notation. For a fermionic field of spin $s$, there are $4\cdot(2s+1)$ modes of equal frequency (the factor of 4 takes into account particles and antiparticles in both regions I and II). To define our Fock basis we must select a specific operator ordering for the creation operators associated to these modes.  A state with a definite number of particles in the Rindler basis will be denoted by $\ket{\alpha_1\ldots\alpha_{4\cdot(2s+1)}}$ where $\alpha_i\in\{0,1\}$ indicates whether the $i$-th mode in the chosen fermionic operator ordering is populated or not. In other words, we can identify each number state by a certain binary number. This notation also applies to any factorisation of the Hilbert space we may perform, as $\mathcal{H}=\mathcal{H}_1\otimes\ldots\mathcal{H}_n$. In this case, a state in $\mathcal{H}$ may be obtained simply by concatenating the binary numbers for states in each $\mathcal{H}_i$.

To calculate the vacuum state and excitations in terms of Rindler modes, we choose the specific operator ordering defined by the fully excited state
\begin{align}\ket{1\ldots1}=\prod_{\sigma}\left(c^\dagger_{\sigma,\text{I}}d^\dagger_{-\sigma,\text{II}}d^\dagger_{\sigma,\text{I}}c^\dagger_{-\sigma,\text{II}}\right)\ket{0}\label{ord2}.\end{align}
Here, $\sigma$ is a label running over the $2s+1$ values of the spin z-component. The ordering \eq{ord2} groups together all the region I operators of a given spin z-component with all the region II operators with the reverse spin z-component. It therefore suggests a factorisation of the Hilbert space as
\begin{align}\mathcal{H}=\bigotimes_{\sigma}\mathcal{H}_\sigma\end{align}
where the vacuum state of each $\mathcal{H}_\sigma$, $\ket{0}_\sigma$, satisfies $C_{\sigma,\text{R}}\ket{0}_\sigma=C_{-\sigma,\text{L}}\ket{0}_\sigma=0$. These relations for any fixed $\sigma$ are exactly those found for the Grassman scalar field which is ubiquitous in the relativistic quantum information literature \cite{AlsingSchul,chapucilla,chapucilla2,Shapoor,Geneferm,Edu9}. Therefore, the problem of finding the vacuum and excitations for arbitrary spin is formally equivalent to $2s+1$ copies of the Grassman scalar case.

We make another factorisation of $\mathcal{H}_\sigma$ into left and right sectors, as is implied by the ordering \eq{ord2} where, for any $\sigma$, the first two operators correspond precisely to the right Unruh mode and the other two correspond to the left Unruh mode. The vacuum for the right sector now obeys the single condition $C_{\sigma,\text{R}}\ket{0}_{\sigma,\text{R}}=0$ and involves only region I particle modes and region II antiparticle modes. Using \eq{umodes}, it is straightforward to verify that
\begin{align}\ket{0}_{\sigma,\text{R}}&=\cos\ro\ket{00}+\sin\ro\ket{11}\nonumber\\&=\left(\cos\ro\mathbf{I}+\sin\ro c^\dagger_{\sigma,\text{I}}d^\dagger_{-\sigma,\text{II}}\right)\ket{0}_\text{Rin}\label{fr}\end{align}
where $\ket{0}_\text{Rin}$ is the Rindler vacuum.

Similarly, for the left sector, one finds
\begin{align}\ket{0}_{\sigma,\text{L}}&=\cos\ro\ket{00}-\sin\ro\ket{11}\nonumber\\&=\left(\cos\ro\mathbf{I}-\sin\ro d^\dagger_{\sigma,\text{I}}c^\dagger_{-\sigma,\text{II}}\right)\ket{0}_\text{Rin}\label{fl}\end{align}
where the extra minus sign comes from the reversed operator ordering (we take the criterion of having region I operators appear before region II operators within a given sector; however, this is purely conventional).

Grouping the results \eq{fr} and \eq{fl} together we find the vacuum for a single $\sigma$ to be
\begin{align}\label{gvac}\ket{0}_\sigma&=\cos^2\ro\ket{0000}-\sin\ro\cos\ro\ket{0011}\nonumber\\&+\sin\ro\cos\ro\ket{1100}-\sin^2\ro\ket{1111}\end{align}
where the notation is implicitly defined by grouping  the operators in \eq{fr} and \eq{fl} as
\begin{align}\ket{1111}=c^\dagger_{\sigma,\text{I}}d^\dagger_{-\sigma,\text{II}}d^\dagger_{\sigma,\text{I}}c^\dagger_{-\sigma,\text{II}}\ket{0}_\text{Rin}.\end{align}
The one-particle excitations are obtained straightforwardly by applying \eq{um} to \eq{gvac},
\begin{align}\label{gexc}\ket{1}_\sigma&=(\qr C^\dagger_{\sigma,\text{R}}+\ql C^\dagger_{\sigma,\text{L}})\ket{0}_\sigma\nonumber\\&=\qr\left[\cos\ro\ket{1000}-\sin\ro\ket{1011}\right]\nonumber\\&+\ql\left[\sin\ro\ket{1101}+\cos\ro\ket{0001}\right].\end{align}
With these, we are nearly done: The vacuum state for a single Unruh mode of arbitrary spin in the operator ordering \eq{ord2} is given by 
\begin{align}\ket{0}_\text{U}=\bigotimes_{\sigma}\ket{0}_\sigma,\label{uvac}\end{align}
where we remind the reader that the tensor product of two states in different spin sectors in our notation is obtained simply by concatenating their expressions. 

In order to compute an arbitrary Unruh excitation of the form
\begin{align}\ket{\sigma_1,\dots,\sigma_N}_\text{U}=C^\dagger_{\sigma_1,\text{U}}\ldots C^\dagger_{\sigma_N,\text{U}}\ket{0}_\text{U}\end{align}
we only have to rearrange the operators $C^\dagger_{\sigma_i,\text{U}}$ so that they have the same ordering as the product in \eq{ord2}, and then substitute the factors $\ket{0}_{\sigma_i}$ by $\ket{1}_{\sigma_i}$ in \eq{uvac}. This is possible because the vacuum states for each sector $\ket{0}_{\sigma}$ are superpositions of terms with an even number of particles and therefore no anticommutation signs appear when the operator $C^\dagger_{\sigma',\text{U}}$ `goes through' the operators in sector $\sigma$. 

Some final considerations are in order. As mentioned above, only Dirac fermions 	have been considered so far. The translation of these results to Majorana fermions is straightforward since, although the distinction between particle and antiparticle modes of the same helicity is lost, the Unruh modes \eq{umodes} mix particles of different helicities. The Majorana case is therefore exactly analogous to that of the Grassman scalar field, with particles of negative helicity playing the role of antiparticles.

Finally, we remark that the state coefficients in the basis related to any other operator ordering different from \eq{ord2} can be readily obtained from the above expressions by simply rearranging the operators. Therefore, the coefficients in any ordering differ from those computed above at most by a sign.

\section{Bosonic fields}\label{bos}
The notation and arguments employed in the previous section for fermionic fields can be carried over to the bosonic case almost without modification. The main differences are that in the bosonic case no sign ambiguity concerning operator ordering may appear, that the number of excitations in each mode is unbounded due to the lack of any Pauli's exclusion principle (and thus the states can no longer be labeled by a binary number), and that in the bosonic case the Unruh modes are given by
\begin{align}\label{umboson}
 A^\dagger_{\omega,\text{R}}&=\cosh r_\omega\, a^\dagger_{\omega,\sigma,\text{I}} - \sinh r_\omega\, a_{\omega,-\sigma,\text{II}},\nonumber\\
 A^\dagger_{\omega,\text{L}}&=\cosh r_\omega\, a^\dagger_{\omega,\sigma,\text{II}} - \sinh r_\omega\, a_{\omega,-\sigma,\text{I}}, \end{align}
 where the parameter $r_\omega$ is now defined by $\tanh r_\omega=e^{-\pi\omega c/a}$. Note that no distinctions are made between particle and antiparticle modes since, contrary to the case of Dirac fermionic fields, antiparticles are not a necessity of the formalism. Should we want to treat a complex field with distinct particles and antiparticles, we would merely add another subscript indicating particle species to the operators. As all the magnitudes that change under time reversal, this label should change in the second term of the Unruh modes \eq{umboson} just like spin does. The Unruh mode under consideration, analogous to \eq{um}, is
 \begin{align}A^\dagger_{\omega,\sigma,\text{U}}=\qr A^\dagger_{\omega,\sigma,\text{R}}+\ql A^\dagger_{\omega,-\sigma,\text{L}}\label{umb}\end{align}
 As in the previous section, we shall henceforth drop the frequency label $\omega$ since it will play no role in our calculations.
 
 As before, we can factor the Hilbert space in a product of the different degrees of freedom of the field
 \begin{align}\mathcal{H}=\bigotimes_{\sigma}\mathcal{H}_\sigma\end{align}
 where $\sigma$ takes $2s+1$ distinct values for a massive field. Although all the operator orderings lead to the same basis in the bosonic case, it is still important to specify the notation we use for the field excitations. We will employ the ordering analogous to \eq{ord2},
 \begin{align}\ket{1\ldots1}=\prod_{\sigma}\left(a^\dagger_{\sigma,\text{I}}a^\dagger_{-\sigma,\text{II}}\right)\ket{0}\label{ord2b}.\end{align}
 The vacuum and arbitrary particle excitations are given, as in the fermionic case, by the expressions 
 \begin{align}\ket{0}_\text{U}=\bigotimes_{\sigma}\ket{0}_\sigma\label{uvacb}\end{align}
  and
\begin{align}\bigotimes_{\sigma}\ket{n_\sigma}_\sigma=\frac{1}{\sqrt{n_1!\ldots n_k!}}(A_{\sigma_1,\text{U}})^{n_1}\ldots (A_{\sigma_k,\text{U}})^{n_k}\ket{0}_\text{U}\label{uexcb}\end{align}
 Notice that the complete state is obtained by concatenating all the different spin sectors.
 
 Therefore, all that remains is to find the vacuum and arbitrary excitation in the Rindler basis for any fixed $\sigma$ subspace. In other words, we only need to compute the vacuum and arbitrary excitation for the scalar field. 
  
Following \cite{Edu9}, we make a squeezed vacuum state ansatz for $\ket{0}_\sigma$
\begin{align}\ket{0}_\sigma=\sum_{n=0}^\infty f(n)\ket{n\, n}\end{align}
where, following our notation, we have
\begin{align}\ket{n\, n}&=\frac{1}{n!}(a^\dagger_{\sigma,\text{I}})^n(a^\dagger_{-\sigma,\text{II}})^n\ket{0}_\text{Rin}.\end{align}
If we now impose the obvious conditions $A_{-\sigma,\text{L}}\ket{0}_\sigma=A_{\sigma,\text{R}}\ket{0}_\sigma=0$ we get the following recurrence relation
\begin{align}\cosh\ro f(n)-\sinh\ro f(n-1)=0\end{align}
with solution $f(n)=C_\text{N}\tanh^n\ro$. The constant $C_\text{N}$ can be found from the normalisation condition
\begin{align}C_\text{N}^2\sum_{n=0}^\infty\tanh^{2n}\ro=1.\end{align}
The geometric series is readily evaluated as
\begin{align} \sum_{n=0}^\infty\tanh^{2n}\ro=\frac{1}{1-\tanh^2\ro}=\cosh^2\ro\end{align}
and therefore $C_\text{N}=1/\cosh\ro$. The vacuum state is then
\begin{align}\ket{0}_\sigma=\frac{1}{\cosh\ro}\sum_{n=0}^\infty \tanh^n\ro\ket{n\, n}\label{ubvac}.\end{align}
 Hence, the one particle excitation is
\begin{align}\ket{1}_\sigma&=(\qr A^\dagger_{\sigma,\text{R}}+\ql A^\dagger_{-\sigma,\text{L}})\ket{0_\text{U}}=\sum_{n=0}^\infty f(n)\frac{\sqrt{n+1}}{\cosh\ro}\ket{\Phi^n},\nonumber\\\ket{\Phi^n}&=\ql\ket{n\, (n+1)}+\qr\ket{(n+1)\, n}.\end{align}

With these results, the higher spin analogs of all the states previously considered in the literature can be readily studied. For higher excitations, a recurrence relation can be found: If we write the excitation as
\begin{align}\ket{n}_\sigma=\sum g_n(k,l)\ket{k\, l},\end{align}
then applying the Unruh creation operator and dividing by $\sqrt{n+1}$ to normalise we obtain the recurrence relations 
\begin{align}g_{n+1}(k+1,l)&=\frac{\qr}{\sqrt{n+1}}\left[\sqrt{k+1}\cosh\ro g(k,l)\right.\nonumber\\&+\left.\sqrt{l+1}\sinh\ro g(k+1,l+1)\right],\nonumber\\
g_{n+1}(k,l+1)&=\frac{\ql}{\sqrt{n+1}}\left[\sqrt{l+1}\cosh\ro g(k,l)\right.\nonumber\\&+\left.\sqrt{k+1}\sinh\ro g(k+1,l+1)\right].\end{align}
These relations, together with the expression \eq{ubvac} for the vacuum state, uniquely determine the arbitrary particle excitations.

\section{Entanglement in fields of higher spin}\label{entsec}
In this section we study entanglement in bipartite field states of arbitrary spin of the form
\begin{align}\ket{\Psi}=\frac{1}{\sqrt{2}}\left(\ket{0}_\text{A}\ket{A}_\text{R}+\ket{0}_\text{A}\ket{B}_\text{R}\right)\label{genst}\end{align}
where $\{\ket{0}_\text{A},\ket{1}_\text{A}\}$ is a qubit Hilbert space basis for Alice, who is customarily taken to be watching an inertial field mode (i.e. Alice is an inertial observer) and $\{\ket{A}_\text{R},\ket{B}_\text{R}\}$ are two states obtained by applying an arbitrary linear combination of products of Unruh creation operators to the Minkowski vacuum, at a frequency very different from Alice's modes so that their overlap is negligible. These states comprise the second part of the system, which is watched by an uniformly accelerated observer (Rob) moving in region I of Minkowski spacetime. Since Rob is non-inertial, the natural coordinates to describe the field from his viewpoint are Rindler coordinates and, thus, their associated Rindler basis. 

Also, since Rob is unable to access the field outside region I, he must trace over region II modes to obtain a physical mixed state which describes the correlations in the Alice-Rob bipartite system.

It is in this reduced state where we will study entanglement. We employ the negativity \cite{Negat}, an entanglement measure suited for the study of mixed states. It is defined as the absolute value of the sum of the negative eigenvalues of the partial transpose matrix.

The results obtained in sections \ref{ferm} and \ref{bos} allow us to express any field state in the Rindler basis and also provide new tools which make the study of entanglement in some settings trivial. For instance, looking at \eq{uvacb} or \eq{uvac} we see that if we have a state in which only a single $\sigma$ is excited, say $\sigma_i$, then the state will factor as
\begin{align}\ket{\Psi}=\ket{\Psi}_{\sigma_i}\otimes\left(\bigotimes_{j\neq i}\ket{0}_{\sigma_j}\right).\label{fact}\end{align}
If the state \eq{fact} is entangled, all of the entanglement must be in the factor $\ket{\Psi}_{\sigma_i}$, which implies that the entanglement in the states $\ket{\Psi}_{\sigma_i}$ and $\ket{\Psi}$ are the same. Thus, the existence of this spin factorisation explains the universality phenomenon found \cite{Edu2,Edu3} where the Grassman field state
\begin{align}\frac{1}{\sqrt{2}}\left(\ket{0}_\text{A}\ket{0}_\text{R}\pm\ket{1}_\text{A}\ket{1}_\text{R}\right)\end{align}
and the Dirac field state
\begin{align} \frac{1}{\sqrt{2}}\left(\ket{0}_\text{A}\ket{0}_\text{R}\pm\ket{1}_\text{A}\ket{\sigma}_\text{R}\right)\end{align}
with $\sigma\in\{\up,\down\}$ were found to have exactly the same entanglement. Notice that this argument requires the use of a Hilbert space basis associated with a specific operator ordering. However, as seen in section \ref{ferm} and studied detailedly in \cite{Mig2}, entanglement changes when different operator orderings are chosen. Nevertheless, it can be shown the equality remains true for any other ordering.

The same arguments hold for bosonic fields even more directly as in this case there is no operator ordering ambiguity.  This means that the massless spin 1 state 
\begin{align}\frac{1}{\sqrt{2}}\left(\ket{0}_\text{A}\ket{0}_\text{R}\pm\ket{1}_\text{A}\ket{p}_\text{R}\right),\label{stats}\end{align}
where $p\in\{L,R\}$ describes helicity, has the same entanglement properties as the scalar field state
\begin{align} \frac{1}{\sqrt{2}}\left(\ket{0}_\text{A}\ket{0}_\text{R}\pm\ket{1}_\text{A}\ket{1}_\text{R}\right).\end{align} 

We now study entanglement in slightly less trivial states, using the results of sections \ref{ferm} and \ref{bos} to express Rob's part of the state in the Rindler basis. We consider both the massless spin 1 case and the massive spin $3/2$ case.
\subsection{Spin one}\label{mcPKing}

This is a very interesting case since it corresponds to the electromagnetic field. Non-inertial entanglement for the electromagnetic field has been examined before in \cite{LingHeZ}. However, several technical misconceptions invalidate those previous results\footnote{Namely, in \cite{LingHeZ} the authors did not consider the correct product of the two different spin sectors that appear for the electromagnetic field. This resulted in a wrong vacuum state, as can be checked by applying annihilator operators to it. As a consequence, this led to the incorrect result that entanglement is not affected by acceleration.}. Here we will see that, as it happens to the fermionic field \cite{Edu2}, bosonic entanglement in the spin degree of freedom is affected by acceleration in a very similar way as occupation number entanglement.

\begin{figure}[hbtp] \centering
\includegraphics[width=.50\textwidth]{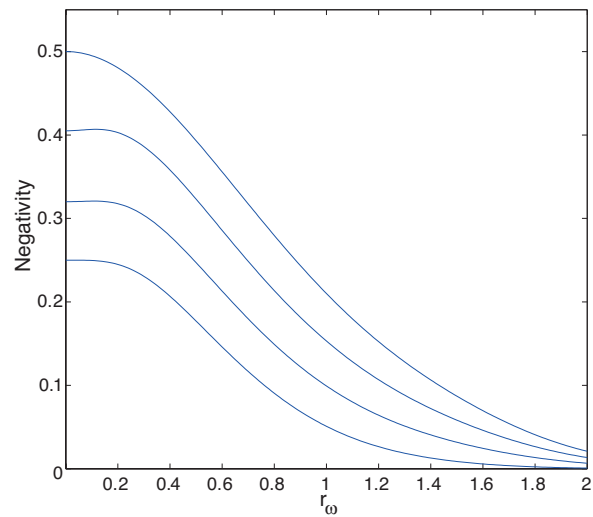}
\caption{Negativity as a function of $r_\omega$ for the state \eq{spin1} and different values of $\qr$. From top to bottom, $\qr=1,0.9,0.8,1/\sqrt{2}$. Note the slight bump for $\qr=0.9$ and small $r_\omega$ which depicts entanglement creation.}
\label{sp1}
\end{figure}

Figure \ref{sp1} shows the negativity as a function of $r_\omega$ and different values of $\qr$ for the massless spin 1 state
\begin{align}\ket{\Psi}_\text{B}=\frac{1}{\sqrt{2}}\left(\ket{R}_\text{A}\ket{L}_\text{R}\pm\ket{L}_\text{A}\ket{R}_\text{R}\right).\label{spin1}\end{align}
The results are qualitatively similar to those found in \cite{Edu9}. Entanglement is completely degraded in the infinite acceleration limit and there is less inertial entanglement in the initial state as $\qr$ increases. However, there is a remarkable difference with the scalar field results reported in \cite{Edu9}: For $\qr=0.9$, Figure \ref{sp1} shows a small increase in entanglement for small $r_\omega$. This is another instance of the entanglement creation phenomenon reported in \cite{Mig1}, where only bosonic scalar and Grassman scalar fields were considered. These results therefore show explicitly that this entanglement creation phenomenon can also happen for formally maximally entangled states such as \eq{stats}.

We would like to remark that in \cite{polacos}, a qualitatively similar phenomenon of an entanglement  maximum in a special relativistic context is reported. However, the similarities are only superficial: Our results present negativity, while \cite{polacos} study Clauser-Horne-Shimony-Holt correlations. We study the behaviour of entanglement under uniform acceleration, and therefore we are forced to trace out modes causally disconnected from the observer. The maximum in fig.\ref{sp1} is the result of two competing trends: On one hand, the change of basis from Minkowski to Unruh modes tends to create entanglement, while on the other, the tracing out of modes tends to wash it out.   \cite{polacos} studies entanglement between two inertial parties. Since no tracing of modes is present, their maximum must have a different origin. Finally, we remark that the maximum in fig. \ref{sp1} has an energy proportional to the acceleration of the observer, while the maximum in \cite{polacos} happens at a fixed energy. For reasonable accelerations, both maxima differ by many orders of magnitude.
\subsection{Spin 3/2}
For the spin $3/2$ case, we have to consider a state with more than one-particle Unruh excitations, since otherwise the state would always have a lower-spin analog. We therefore consider the state \eq{genst} with
\begin{align}\ket{A}&=\frac{1}{\sqrt{2}}\left(\ket{\up\nearrow}+\ket{\up}\right),\nonumber\\\ket{B}&=\frac{1}{\sqrt{2}}\left(\ket{\down\searrow}+\ket{\down}\right)\label{spin32}\end{align}
where we have set up the notation
\begin{align}
&\ket{\up}=\ket{S=3/2,\sigma=3/2},\\*
&\ket{\nearrow}=\ket{S=3/2,\sigma=1/2},\\*
&\ket{\searrow}=\ket{S=3/2,\sigma=-1/2},\\*
&\ket{\down}=\ket{S=3/2,\sigma=-3/2},
\end{align}
for the four spin z-component states of the field ($\sigma$).

\begin{figure}[hbtp] \centering
\includegraphics[width=.50\textwidth]{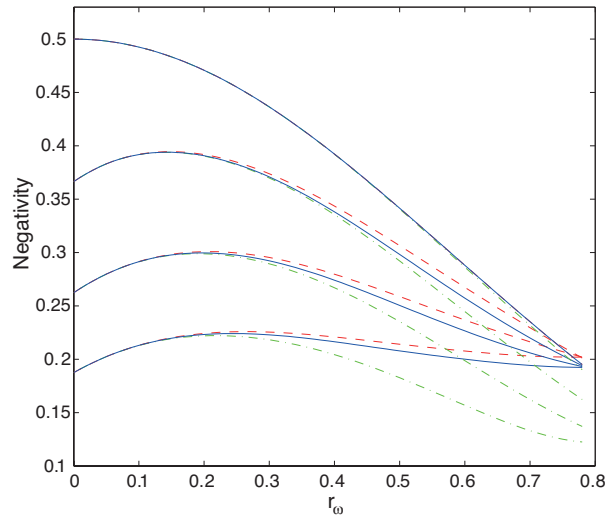}
\caption{(Color online) Negativity as a function of $r_\omega$ for the state \eq{spin32} and different values of $\qr$. From top to bottom, $\qr=1,0.9,0.8,1/\sqrt{2}$. Blue continuous curves show negativity in the `physical' ordering in which all region I operators appear to the left of all region II operators. Red dashed curves correspond to the canonical ordering employed in previous literature \cite{Edu9}. Green dash-dotted curves correspond to negativity in the `spin' operator ordering \eq{ord2}.}
\label{sp32}
\end{figure}

As mentioned before, because of the operator ordering ambiguity present in fermionic fields negativity is not uniquely defined. Figure \ref{sp32} shows the negativity for the state \eq{spin32} and the bases associated with three different operator orderings:

\begin{enumerate}
\item The `spin' ordering \eq{ord2} used in section \ref{ferm} to obtain the expressions for the excitations in arbitrary spins. 
\item  A generalisation of the `canonical' ordering employed in \cite{Edu9}, which exploits the tensor product structure of the whole space in terms of left and right sectors, rather than the spin structure.
\item The physically preferred class of operator orderings discussed in \cite{Mig2}, namely, those orderings in which all region I operators appear to the left of all region II operators. All these orderings result in the same negativity. The last curve in Figure \ref{sp32} represents this `physical' negativity class.
\end{enumerate}

Note that the `physical' and `canonical' negativities lie very close to each other for all values of $\qr$; this is but a quirk of the state \eq{spin32} and does not happen in general. The `spin' ordering in this case happens to deviate significantly from the other two curves. Nevertheless, all three curves present a qualitatively similar behaviour: A maximum entanglement is reached and afterwards it is degraded up to a finite limit, a characteristic which is the hallmark of fermionic statistics \cite{AlsingSchul}. This finite limit is independent of $\qr$ for both the `physical' and the `canonical' negativities, but not so for the `spin' one.

\section{Conclusions}\label{cojoniak}

We have found expressions for the vacuum and Unruh excitations beyond the single mode approximation for fields of arbitrary spin. By taking advantage of an appropriate tensor product structure of the Hilbert space, the problem was reduced to computing these quantities for spin 0, a case well known in the literature.

 The expressions derived here therefore make it straightforward to extend all the previous studies in quantum information to fields of arbitrary spin, both under and beyond the single mode approximation. The formalism developed here can be also used to study other internal degrees of freedom that were not affected by the kinematical state of the observer.

We have applied our formalism to study the most accessible quantum field for performing quantum information, the electromagnetic field, which is of spin 1. Some entanglement amplification was found in the spin 1 singlet state for some values of $\qr\neq1,1/\sqrt{2}$. 

We also considered a representative state for the spin $3/2$ field. We studied the negativities in the bases associated to three different operator orderings: the `spin'  ordering used in section \ref{ferm} to easily compute the vacuum and excitations, the `canonical' ordering used in previous literature \cite{Edu9} and the `physical' ordering as developed in \cite{Mig2}. The entanglement behaviour was qualitatively similar in all these cases.

All our considerations can be of course exported to a setting consisting of two observers in the vicinity of a black hole, one standing still close to the horizon and the other free-falling. The details of this correspondence can be found in \cite{Edu6}.

These results, along with the banishment of the single mode approximation in \cite{Edu9}, provide a fully general formalism to analyse the entanglement of  quantum fields in non-inertial frames.

\section{Acknowledgments}

We thank Rob Mann for his helpful comments. E. M-M was supported by a CSIC JAE-PREDOC2007 Grant and by the Spanish MICINN Project FIS2008-05705/FIS and the QUITEMAD consortium.

\end{document}